\documentclass{article}
\makeatletter

\@addtoreset{equation}{section}
\makeatother
\usepackage{epsf}
\begin{document}
\title{ A Unitarized Chiral Approach to $f_0(980)$ and $a_0(980)$ 
States\\
and Nature of Light Scalar Resonances}
\author{Masayuki \sc Uehara\footnote{E-mail: ueharam@cc.saga-u.ac.jp}\\
Takagise-Nishi 2-10-17, Saga 840-0921, Japan}
\date{\today}
\maketitle
\begin{abstract}
We show that the $f_0(980)$ and $a_0(980)$ states appearing in 
scattering and production processes can be described quite well by 
the two-channel Oller-Oset-Pel\'aez~(OOP) version of 
unitarized chiral theories. It is impossible, however, to deduce a 
decisive conclusion on the nature of them from the fitting to 
experimental data alone.  
Using explicit $N_c$ dependence of parameters in OOP amplitudes we 
demonstrate that light scalar resonances are not of $q\bar q$ states but 
dynamical rescattering effects generated under chiral symmetry, 
unitarity and channel couplings.
 \end{abstract}
\def\beq{\begin{equation}}　\def\eeq{\end{equation}}
\def\beqa{\begin{eqnarray}}　\def\eeqa{\end{eqnarray}}
\def\beqan{\begin{eqnarray*}}　\def\eeqan{\end{eqnarray*}}
\def\ba{\begin{array}}　\def\ea{\end{array}}
\def\noeq{\nonumber}
\def\mpi{m_\pi} \def\fp{f_\pi}　\def\mK{m_K} \def\fK{f_K}
\def\me{m_\eta} \def\fe{f_\eta}　\def\half{\frac{1}{2}}　
\def\der{\partial}
\def\vI{\mbox{\boldmath$I$}}  \def\vJ{\mbox{\boldmath$J$}}
\def\vg{\mbox{\boldmath$g$}}　\def\vt{\mbox{\boldmath$t$}}
\def\vG{\mbox{\boldmath$G$}}　\def\vK{\mbox{\boldmath$K$}}
\def\vS{\mbox{\boldmath$S$}}   \def\vT{\mbox{\boldmath$T$}}
\def\vrho{\mbox{\boldmath$\rho$}} \def\vJ{\mbox{\boldmath$J$}}
\def\del{\delta}
\def\im{{\rm IM}} \def\re{{\rm R}}
\section{Introduction}
 Probably we have not yet reached a  
consensus on the nature of light scalar mesons below 1 GeV\cite{Close,Ochs}. 
Main issues around them are whether they are real propagating particles or 
are dynamical rescattering effects, whether they are of $q\bar q$ states, 
four quark states or $K\bar K$ molecule states, and whether they form a 
nonet within themselves or only one or two are needed to form a different 
nonet with higher mass scalar states. 

In order to study these issues many authors focus on the analyses of the 
$f_0(980)$ and $a_0(980)$ states appearing in decay 
processes such as $\phi\to\gamma f_0(980)$ and $\phi\to\gamma a_0(980)$. 
In most theoretical approaches the radiative $\phi$ decays take place 
through a radiative charged kaon loop diagram so as to satisfy the 
Okubo-Zweig-Iizuka~(OZI) rule. 
The difference in various approaches comes from the way how to construct 
amplitudes $K^+ K^-\to m_1m_2$ including the $f_0(980)$ or $a_0(980)$ 
states, where $m_1$ and $m_2$ are final two pseudoscalar mesons. 
Among the ways we quote models using $f_0(980)$ and $a_0(980)$ 
propagators\cite{AchaIva,AchaGub,Kumano},  unitarized chiral 
approaches\cite{Oller1,Marko,Oller2,Oset,MU06},  a model using 
the linear sigma model\cite{Bramon1,Bramon2}, a compound model 
of chiral $O(p^2)$ terms and propagators\cite{Gokalp}, and a model with 
channel couplings between hadronic and quark channels\cite{Markushin}.  
There are approaches without using the explicit kaon-loop mechanism; 
a model in which production amplitudes are assumed to be proportional to 
scattering amplitudes with real coefficient functions in order to satisfy 
unitarity\cite{BogPen}, 
and a quark model with a radiative quark loop by assuming that 
all these mesons are to be dominantly $q\bar q$ states\cite{Anisovich}. 
Similarly, $\gamma\gamma\to f_0\mbox{  and  }a_0$ processes are 
expected to reveal the quark structure of them\cite{AchaShes}.
How unambiguously can we determine both theoretical and experimental 
branching fraction $B(\phi\to\gamma f_0)$ ?  Calculating a pole and its 
residues of the $f_0(980)$ state, Boglione and Pennington\cite{BogPen} 
have given $B(\phi\to\gamma f_0)$ much smaller than the 
usually quoted value $(3.4\pm 0.4)\times 10^{-4}$\cite{PDG}. 
This indicates that we should be cautious of extracting the correct signal 
of the $f_0(980)$ state from experimental mass distributions.  
If this is the case, it becomes obscure whether we can reach a decisive 
conclusion on the nature of the $f_0(980)$ and $a_0(980)$ states through 
these analyses.

In this paper we attempt to describe the $f_0(980)$ and $a_0(980)$ states 
appearing in scattering and production processes within the two-channel 
Oller-Oset-Pel\'aez version\cite{OOP}, that is an approximate version of 
Inverse Amplitude Method~(IAM)\cite{ DP,Hanna,GO,GNP,PGN} to unitarize 
amplitudes up to $O(p^4)$ of Chiral Perturbation 
Theory~(ChPT)\cite{GL1,GL2}. 
The basic ingredients of SU(3)$\times$SU(3) ChPT are the octet 
Nambu-Goldstone particles $\pi,~K,~\eta_8$, where we treat $\eta_8$ as the 
physical $\eta$ in the two-channel OOP version, and any scalar 
meson fields are not introduced as independent degrees of 
freedom in advance. It offers, therefore, a good theoretical framework to 
study the issues. Of course, this does not imply that the existence or 
non-existence of a resonance can be predicted by IAM, since ChPT 
amplitudes of $O(p^4)$ include a set of phenomenological parameters, 
low energy constants~(LECs), which are to be  determined so as to reproduce  
experimental data. After having fixed the LECs, the multi-channel 
amplitudes are determined without any additional free parameters.
Though our set of the LECs are not the best solution, the phase shifts are 
reproduced quite well. Since two-meson mass spectra of the radiative $\phi$ 
meson decays depend on the off-diagonal $K^+K^-\to \pi\pi$ and 
$K^+K^-\to\pi\eta$ scattering amplitudes below the $K\bar K$ threshold, 
the resulting mass spectra offer a proving ground of the OOP amplitudes. 
It turns out that the mass spectra are reproduced fairly well by the OOP 
amplitudes. In $\gamma\gamma\to \pi\pi$ processes, however, the 
$f_0(980)$ signal is masked by complicated interferences, since 
various production mechanism contribute coherently. 
This process is a proving ground of production mechanisms, if we have more 
accurate data near the $f_0$ state, therefore. 

Even if IAM or its OOP version can give simultaneous fits to 
scattering and production data, it is impossible to deduce a decisive 
conclusion on the quark structure of the $f_0(980)$ and $a_0(980)$ 
states because of the reason stated above on the LECs. 
We should note, however, that the pion decay constant $f_\pi$ and the LECs 
in ChPT have a definite $N_c$ dependence, where $N_c$ is the number of 
colors. Due to this nature of the theory we can study how scalar amplitudes 
behave when $N_c$ increases from 3. It has been shown recently that the 
scalar states do not survive as $N_c$ is larger than 3, and then the scalar 
states are probably not $q\bar q$ states but dynamical ones, while vector 
mesons survive as narrow resonances as typical $q\bar q$ 
states\cite{Pelaez, MU0801}.  
In this sense IAM and its approximate OOP version can give not only a 
unified description of scattering and production processes but also an 
insight into the quark structure of the light scalar mesons.

 We discuss the OOP description of scattering processes in the next section,  
 including a discussion on LECs used in this 
 paper. Scalar meson productions in radiative  $\phi$ meson decays and 
 $\gamma\gamma$ collisions are discussed in Sec. {\bf 3},  and the large 
 $N_c$ behavior of scalar states are discussed in Sec.{\bf 4}, and  
 concluding remarks are given in the last section.

\section{The OOP description of scattering processes}
\subsection{The OOP amplitude}
The ingredients of IAM consist of chiral order $O(p^2)$ and $O(p^4)$ 
amplitudes of ChPT. An $O(p^2)$ amplitude, denoted by  $T^{(2)}(s,t,u)$, 
has a form of a linear function of $s,~t,~u$ and meson mass squared divided 
by $f_\pi^2$, where $s,~t$ and $u$ are the usual Mandelstam variables. 
An $O(p^4)$ amplitude is composed of polynomial terms and 
one-loop terms: A polynomial term, denoted by $T^{(4)}_{\rm poly}(s,t,u)$, 
is given as a sum of quadratic functions of $s,~t,~u$ and meson mass squared 
with the LECs, denoted as $L_n$, as follows:
\beq
T^{(4)}_{\rm poly}(s,t,u)=\sum_{n=1,8}\frac{1}{f_\pi^4}L_n~P_n(s,t,u),
\eeq
where $P_n$  are the quadratic functions. Loop terms are given by  
$T^{(2)}\times T^{(2)}$ with divergent loop integrals, which are 
regularized in the $\overline{MS}-1$ scheme at the renormalization scale 
$\mu$\cite{GL1,GL2}. An $s$-channel loop term is given by 
$t^{(2)}(s)J(s)t^{(2)}(s)$, where $t^{(2)}$ is a partial wave amplitude 
derived from $T^{(2)}(s,t,u)$ and $J(s)$ is the one-loop function which 
will be given soon. There are similar $t$- and $u$-channel loop terms 
and tadpole terms, but they are discarded in the OOP version. 
The ingredients of amplitudes in the OOP version consist of the $O(p^2)$ 
terms, the polynomial and the $s$-channel loop terms of $O(p^4)$ amplitudes. 
We use the $O(p^4)$ amplitudes given in Ref.~\cite{GNP}, where the kaon 
and $\eta$ decay constants are set equal to the pion decay constant 
$f_\pi$ in order to guarantee exact perturbative unitarity.

The two-channel OOP amplitudes are written in the symmetric $2\times2$ 
matrix form as 
\beq
\vT(s)=\vt^{(2)}(s)[\vt^{(2)}(s)-\vt^{(4)}(s)]^{-1}\vt^{(2)}(s),
\eeq
where $\vt^{(2)}(\vt^{(4)})$ is the $2\times2$ partial wave amplitude 
with chiral order $O(p^2)$~($O(p^4)$) amplitude and $s$ is the total 
center of mass~(CM) energy squared.  Our T-matrix is normalized as  
\beqa
S_{ij}(s)&=&\del_{ij}-2i\rho_i^{1/2}(s)T_{ij}(s)\rho_j^{1/2}(s),\\
\rho_i(s)&=&\frac{1}{16\pi}\frac{2k_i}{\sqrt{s}}\theta(s-s_i)
\eeqa
where $\rho_i$ is the phase space factor with $k_i~(s_i)$ being the 
CM momentum~(threshold energy squared) of the $i$-th channel.
The partial wave $\vt^{(4)}(s)$ is given as
\beq
\vt^{(4)}(s)=\vt^{(4)}_{\rm poly}(s)+\vt^{(2)}(s)\cdot\vJ(s)\cdot
\vt^{(2)}(s), \label{OOPamp}
\eeq
where $\vt^{(4)}_{\rm poly}(s)$ is the partial waves of  
$T^{(4)}_{\rm poly}(s,t,u)$, and 
$\vJ(s)$ is the diagonal $2\times 2$ loop integral; $J_i(s)$ of the 
$i$-th channel is written  
\beqa
J_i(s)&=&\frac{1}{(4\pi)^2}\left\{-1+
\log\left(\frac{m_1m_2}{\mu^2}\right)
+\frac{\Delta_{21}}{s}\log\left(\frac{m_2}{m_1}\right)\right.
 \noeq\\
&+&\left. \lambda_i(s)\log\left(
\frac{\sigma_{i+}(s)+\sigma_{i-}(s)}
{\sigma_{i+}(s)-\sigma_{i-}(s)}\right)\right\}, \label{Loop}\\
\sigma_{i\pm}&=&\sqrt{1-(m_1\pm m_2)^2/s},\\
\lambda_i(s)&=&\sigma_{i+}(s)\sigma_{i-}(s)
\eeqa
for the channel with unequal masses $m_1$ and $m_2$ in the $i$-th channel, 
and $\Delta_{21}=m_2^2-m_1^2$. Because the imaginary part of $J_i(s)$ is 
given as 
\beq
{\rm Im}J_i(s)=-\rho_i(s),
\eeq
$\vT(s)$ satisfies the exact $s$-channel unitarity relation 
\beq
{\rm Im}T_{ij}(s)=-T_{ik}^*(s)\rho_k (s)T_{kj}(s).
\eeq

The LECs, $L_1$ to $L_8$, are determined so as to reproduce $S$-wave 
$\pi\pi$ phase shifts up to about 1.2 GeV. It should be noted that our 
fitting covers a wide energy range 
from the $\pi\pi$ threshold to 1 GeV or more including resonances, that 
the LECs appear in $\vT$ non-linearly and that the OOP version discards 
the $t$- and $u$-channel loop terms and tadpole terms of the full $O(p^4)$ 
amplitudes. So that our set of the LECs and $\mu$ could be different 
from those of ChPT determined through low energy data below the lowest 
resonance region. The scale $\mu$ has a meaning corresponding to a 
cut-off parameter as discussed in Ref.~\cite{OOP}, and we set 
$\mu=0.87$ GeV, that is larger than usually adopted value $\mu=m_\rho$ 
by about 100 MeV. 
Our set of the LECs are given as
\beq
\begin{array}{ccc}
L_1=0.70\times 10^{-3},& L_2=1.30\times 10^{-3},&L_3=-3.20
\times 10^{-3},\\
L_5= 1.50\times 10^{-3},&L_7=-0.25\times 10^{-3},&L_8=0.71\times 10^{-3},
\end{array}\label{LECs}
\eeq
and $L_4$ and $L_6$ are fixed equal to zero.

\subsection{The $(\vI,~\vJ)=(0,~0)$ channel: $\pi\pi\times K\bar K$}
The characteristic behavior of this channel is that the $\pi\pi$ phase 
shift $\delta^{00}_{11}$ rises from the $\pi\pi$ threshold, forms a 
plateau of $50^\circ\sim 80^\circ$ from 500 to 800 MeV, 
crosses $90^\circ$ in the region $800\sim 900$ MeV, and suddenly 
increases over $200^\circ$ just below the $K\bar K$ threshold. 
This feature is reproduced quite well as shown in 
Fig.~\ref{fig:fig00}~(a). The phase shift calculated in the single 
$\pi\pi$ channel formalism cannot exceed $90^\circ$ as shown by the dotted 
line in Fig.~{\ref{fig:fig00}~(a). On the other hand a bound state pole 
appears in the $K\bar K$ channel, when the channel coupling to the 
$\pi\pi$ channel is switched off.  A weak coupling to the $\pi\pi$ channel 
generates a narrow resonance behavior in the $\pi\pi$ channel. 
The steep rising of the $\pi\pi$ phase shift near the $K\bar K$ threshold 
is due to the interference between the $\pi\pi$ background and the 
narrow resonance born as the $K\bar K$ bound state. In this sense the 
$f_0(980)$ state is called a bound state 
resonance\cite{Wein,Lohse,Janssen,OOBS, MU04}. Also shown 
are the production rate defined by $(1-\eta_{00}^2)/4 $ with 
$\eta_{00}$ being the inelasticity and the the phase, 
$\delta^{00}_{12}=\delta^{00}_{11}+\delta^{00}_{22}$, above the 
$K\bar K$ threshold. These give the size and the phase of 
the off-diagonal scattering amplitude $T_{12}$ above the $K\bar K$ 
threshold. 

In our set of the LECs and $\mu$ the phase shift $\delta^{00}_{11}$ 
crosses $90^\circ$ at 875 MeV. The scattering lengths are given as 
\beq
a_{00}=0.217~m_\pi^{-1}\qquad a_{20}=-0.041~m_\pi^{-1}. 
\eeq
According to Ref.~\cite{CGL} ChPT up to two loop accuracy gives 
\beq
a_{00}=0.220\pm 0.005\qquad a_{20}=-0.0444\pm 0.0010. 
\eeq  
\begin{center}
\begin{figure}[h]
\epsfxsize=12 cm
\centerline{\epsfbox{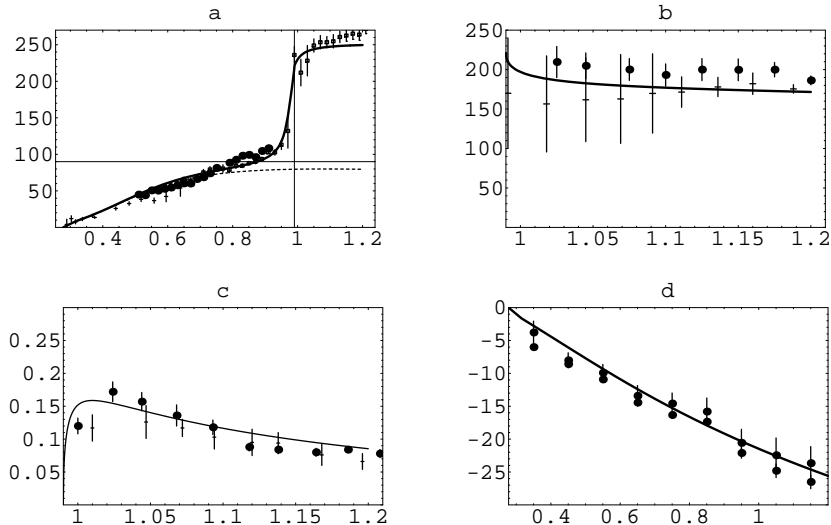}}
\caption{(a) The $\pi\pi$ phase shift $\delta^{00}_{11}$. The solid~(dotted) 
line represents the result of the two-~(single $\pi\pi$-)channel   
calculation. The vertical~(horizontal) line represents the $K\bar K$ 
threshold~($90^\circ$). Experimental data are full 
circles~\cite{estamartin}, blank squares~\cite{CERNb}, 
bars~\cite{sriniv,rossel}, and triangles~\cite{prot}. 
(b) $\pi\pi\to K\bar K$ phase, $\delta^{00}_{12}$. Experimental data 
are taken from \cite{cohen}.  
(c) $(1-\eta_{00}^2)/4$. The data are full circles\cite{cohen} and 
bars\cite{martin}. (d)  $\delta^{20}$ with the data from \cite{hoog}.  
The abscissas are the invariant $\pi\pi$ mass in units of GeV.}
\label{fig:fig00}
\end{figure} 
\end{center}
The phase difference $\delta^{(00)}-\delta^{(20)}$ at the kaon mass is 
$49.9^\circ$, which is the phase of two-pion decay amplitudes of a 
kaon originating from the final state interactions. Recently the KLOE 
collaboration gives $(48\pm 3)^\circ$\cite{KLfsi}, and ChPT does   
$(47.7\pm 1.5)^\circ$\cite{CGL}. 
\begin{center}
\begin{figure}[h]
\epsfxsize=14 cm
\centerline{\epsfbox{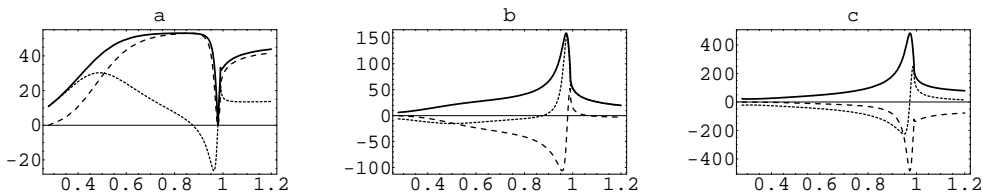}}
\caption{ Energy dependence of $-T_{ij}$. (a) $-T_{11}$. (b) $-T_{12}$. 
(c) $-T_{22}$. The solid lines are the absolute values, the dotted lines  
the real parts, and the dashed lines the imaginary parts.}
\label{fig:f0amp}
 \end{figure} 
\end{center}

We also show all of the amplitudes $T_{ij}$ in Fig.~\ref{fig:f0amp}.  
The $f_0$ structure appears as a sharp dip in $T_{11}$ owing 
to the interference with the large background, which is often called 
the $\sigma(600)$ state, but $T_{12}$ and $T_{22}$ show the clear peak 
corresponding to the $f_0$ state just below the $K\bar K$ threshold.   
The magnitude and shape of $T_{12}$ below the $K\bar K$ threshold is  
crucial to the $\phi\to \gamma \pi\pi$ decays, and it turns out that  
our OOP amplitude $T_{12}$ reproduces the $\pi\pi$ mass spectra fairly 
well.

\subsection{The $(\vI,~\vJ)=(1,~0)$ channel: $\pi\eta\times K\bar K$}
Since there are almost no experimental data on extrapolated 
$\pi\eta$ elastic scattering, the $\pi\eta$ mass distributions from 
$K^-p\to\pi^-\eta\Sigma^+(1385)$ reaction\cite{amsterdam} and 
$pp\to p(\eta\pi^+\pi^-)p$ reaction\cite{wa76} have been compared 
with theoretical works\cite{GNP,OOP,OOBS,OOND}. The former reaction 
is expected to give data of the $K\bar K\to\pi\eta$ process, and the 
latter to derive information on the elastic $\pi\eta\to\pi\eta$ process, 
but the data is not sufficient. We do not give the phase shift and cross 
section here, therefore. (These are shown in Sec.{\bf 4}.) 

The $(K\bar K)_{I=1}$ scattering amplitude in the single channel 
formalism has a bound state, but the situation is much different from 
the $f_0$ case; the bound state disappears when the $\pi\eta$ coupling is 
switched off, and the $\pi\eta$ elastic channel is repulsive in the single 
channel formalism. The channel coupling between the $\pi\eta$ and 
$K\bar K$ channels is strong in this channel, however, and then the 
production rate $(1-\eta_{(10)}^2)/4$ reaches almost the maximum value 
$0.25$. This strong channel coupling generates the $a_0$ structure as 
pointed out in the meson-exchange model\cite{Janssen} and unitarized chiral 
theories\cite{OOBS,MU04}. Thus, the generating mechanism of $a_0(980)$ 
is different from that of $f_0(980)$; the channel coupling is crucial 
for both, but it is strong for the former but weak for the latter.   
\begin{center}
\begin{figure}[h]
\epsfxsize=14cm
\centerline{\epsfbox{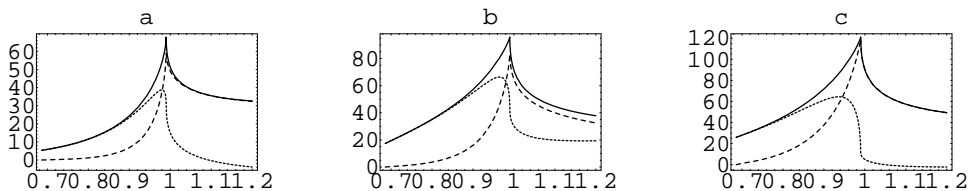}}
\caption{Energy dependence of $-T_{ij}$. (a) $-T_{11}$. (b) $-T_{12}$. 
(c) $-T_{22}$.  The lines are the same as in Fig.2.}
\label{fig:amp10}
 \end{figure}
 \end{center}

The amplitudes $T_{ij}$ in the $(1,0)$ channel show the common shape 
having a sharp cusp-like peak at the $K\bar K$ threshold as shown in 
Fig.~\ref{fig:amp10}. This feature will be 
examined in the $\pi\eta$ mass spectra observed in radiative reactions
$\phi\to\pi\eta$ and $\gamma\gamma\to\pi\eta$. 

\subsection{Other partial waves}
The $(I,~J)=(1/2,0)$ and $(3/2)$ phase shifts are 
described qualitatively well by our OOP amplitudes, though the results 
are not shown. But better fitting is obtained if we use $f_K=1.22f_\pi$. 
 The scattering lengths are 
\beq
a_{1/2,0}=0.216~(0.161)\qquad a_{3/2,0}=-0.054~(-0.049)
\eeq 
in units of $(m_\pi^{-1})$, where the values in the parentheses are for 
the case that $f_K=1.22f_\pi$, while ChPT gives \cite{BKM}
\beq
a_{1/2,0}=0.17\pm 0.02\qquad a_{3/2,0}=-0.05\pm 0.02.
\eeq

The vector resonances, $\rho$ and $K^*(890)$, are also reproduced well 
by our OOP amplitudes. The mass and width of the $\rho$ meson 
are 784 MeV and 162 Mev, respectively.  Those of the $K^*$ meson are 
818 MeV and 28 MeV, respectively, for $f_K=f_\pi$.  In the case that  
$f_K=1.22f_\pi$ the obtained values are 900 MeV and 45 MeV, respectively. 

Thus, we say that the validity of the two-channel OOP amplitudes is 
confirmed.

 \subsection{Pole search for the $f_0$ and $a_0$ states}
The OOP amplitudes $T_{ij}$ can be decomposed into a pole term and its 
background term as
\beq
T_{ij}(s)=\Gamma_i(s)\Delta(s)\Gamma_j(s)+U_{ij}(s), \label{Ida}
\eeq
where $\Gamma_i(s)$ is the one-particle irreducible form factor of the 
$i$-th channel, $\Delta(s)$ is the propagator of the relevant particle, 
and $U_{ij}$ is the background amplitude satisfying unitarity by 
themselves\cite{Ida}. The form factor $\Gamma_i$ satisfies the unitarity 
relation,
\beq
{\rm Im}\Gamma_i(s)=-U^*_{ij}(s)\rho_j(s)\Gamma_j(s), \label{FF}
\eeq
and then the phase of $\Gamma_i$ is controlled by those of $U_{ij}$. 

The propagator $\Delta(s)$ continued to the second Riemann sheet has a 
complex conjugate pair of poles, and the pole near the real axis on 
the lower half-plain contributes to the physical amplitude  as 
\beq
P_{ij}=
g_ie^{i\alpha_i}\left[\frac{1}{s-W_R^2}\right] g_je^{i\alpha_j},
\eeq
where  $g_ie^{\alpha_i}$ is the form factor $\Gamma_i(s)$ at 
$s=M_R^2$, and $W_R=M_R-iM_I$ with $M_R~(M_I)$ being the 
real~(imaginary) part of the pole. The poles and residues are tabulated in 
Table~\ref{tab:pole}.
 \begin{table}[h]
 \begin{center}
 \begin{tabular}{|c|c||c|c|}\hline
\multicolumn{2}{|c||}{$f_0(980)$} &\multicolumn{2}{c|}{$a_0(980)$}\\ \hline
\multicolumn{2}{|c||}{$W_R=(978.4-19.7\,i)$~MeV} &
\multicolumn{2}{c|}{$W_R=( 1110.4-8.9\,i)$~MeV}\\\hline
$g_1=1.54$~(GeV)&$g_2=4.57$~GeV&$g_1=6.63$~GeV&$g_2=8.50$~GeV\\ \hline
$\alpha_1= -85.6^\circ$&$\alpha_2=-15.8^\circ$ &$\alpha_1=-8.5^\circ$ &
$\alpha_2=-36.1^\circ$\\\hline
\end{tabular}
\caption{Pole positions, residues and phases of the $f_0(980)$ and 
$a_0(980)$ states.}
\label{tab:pole}
\end{center}
\end{table}

It should be noticed that the pole of the $a_0$ state appears above the 
$K\bar K$ threshold by 120 MeV, while the pole of the $f_0$ state exists 
below the threshold and the distance between the two poles is about 130 
MeV. Nevertheless all the amplitudes $T_{ij}$ of the isovector channel 
show the cusp-like behavior at the $K\bar K$ threshold, and then the 
two peaks appear very close to each other in physical processes.  
The large phase $\alpha_1$ for the $f_0$ pole indicates a 
large contribution from the background of the $\pi\pi\to\pi\pi$ 
amplitude, that is due to the $\sigma$ enhancement centered at 500 MeV. 
On the other hand $\alpha_1$ for the $a_0$ pole is small, and then 
$\pi\eta\to\pi\eta$ scattering may 
be dominated by the pole term, but the size of $\alpha_2$ indicate that  
background contribution is not ignored. 

We quote examples of the poles and their residues of previous works:  
\beq
\begin{array}{rcl}
f_0\quad W_R&=&(0.987-0.011\,i)~{\rm GeV} \quad 
(g_1,~g_2)=(1.18,~3.83)~{\rm GeV},\\
a_0\quad W_R&=&(1.030-0.086\,i)~{\rm GeV} \quad 
(g_1,~g_2)=(4.08,~5.60)~{\rm GeV}
 \end{array} \label{Oller}
\eeq
in the two-channel calculation\cite{Oller3}, and 
\beq
f_0\quad W_R=(0.989-0.022~i)~{\rm GeV}\quad
(g_1,~g_2)=(1.155,~1.227)~{\rm GeV}
 \label{BogPen} 
\eeq
in \cite{BogPen}, where $g_2$ is much smaller than that of 
Eq.(\ref{Oller}) and our value in Table 2.  

If we define the partial width as an integral\cite{BogPen},
\beq
\Gamma_i=\frac{2M_I}{\pi}\int_{s_i}^\infty ds
\frac{\rho_i(s)g_i^2}{|W_R^2-s|^2}, \label{integwidth}
\eeq
we have 
\beq
\Gamma_{\pi\pi}=45.3 ~{\rm MeV}\quad \Gamma_{K\bar K}=39.4~{\rm MeV}
\label{f0pole}
\eeq
for the $f_0(980)$ state, and the sum of the partial widths, $77.8$ MeV, 
is larger than $2M_I=35.4$ MeV by a factor 2.  
As to the $a_0(980)$ state the same definition gives huge partial widths; 
\beq
\Gamma_{\pi\eta}=~567{\rm MeV}\quad \Gamma_{K\bar K}=554~{\rm MeV}
\eeq
because of the large residues. Such huge partial widths have already 
been pointed out by Flatt\'e\cite{Flatte}, who shows that the partial 
$\pi\eta$ width could be 300 MeV, when the Flatt\'e phase space factor 
$\rho_F(s)$ is usued, where $\rho_F(s)=\rho_2(s)$ for $s>s_2$, and 
$\rho_F(s)=i|\rho_2(s)|$ for $s<s_2$. These facts cast doubt 
not only on the above definition of the partial widths but also the 
conventional resonance interpretation of $a_0$ in contrast to an isolated 
narrow resonance. 

In order to to extract the pole contribution from the whole amplitudes, 
we have to express the pole term $P_{ij}(s)$ in a region away from the 
pole position. After testing  the Breit-Wigner form,  the Flatt\'e 
form and  the loop form used in Ref.~\cite{AchaGub}, 
we find that the Breit-Wigner form is suitable to the $f_0$ state, where  
we compare the Breit-Wigner amplitudes 
\beq
P_{ij}(s)=\frac{g_ig_j}{s-M_R^2+iM_R(\rho_1g_1^2+\rho_2g_2^2)}
\eeq
with the $T_{ij}$ amplitude; we observe that the two $12$ amplitudes 
resemble with each other near the $f_0$ resonance peak, except for 
the phase because of neglecting the phase $\alpha_{12}\sim 90^\circ$ in 
$P_{12}$. See Fig.~\ref{fig:comp}.
\begin{center}
\begin{figure}[h]
\epsfxsize=10 cm
\centerline{\epsfbox{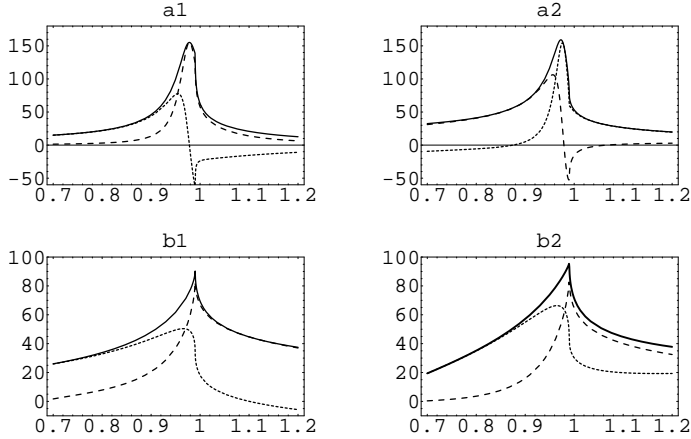}}
\caption{(a)  The $f_0$ state. The left side is $-P_{12}$ and the right 
side $+{\rm Re}T_{12}$, $-{\rm Im}T_{12}$. (b) The $a_0$ state. The left 
side is $-P_{12}$ and the right 
side is $-T_{12}$. The solid lines are the absolute values, 
the dotted lines the real parts and the dashed lines the imaginary parts. 
Note that the real and imaginary parts are exchanged between the left and 
right side of (a) because of $\alpha_1+\alpha_2\sim -90^\circ$. } 
\label{fig:comp}
\end{figure} 
\end{center}
But it turns out that in order to reproduce the sharp cusp-like behavior 
in the isovector channel the Flatt\'e form is required, where $\rho_F$ is 
used instead of $\rho_2$, as shown in Fig.~\ref{fig:comp}~(b). 
The loop form is useful to the $f_0$ state, but not for $a_0$ state, 
because the depression of the real part is insufficient below the 
$K\bar K$ threshold. The residues strongly depress scattering 
amplitudes below the $K\bar K$ threshold in the Flatt\'e form.

Another method to extract the pure $f_0$ amplitude is to assume the form 
of $U_{ij}$ as a single channel elastic amplitude as
\beq
 U_{ij}(s)=\frac{[t^{(2)}_{11}(s)]^2}{t^{(2)}_{11}(s)-t^{(4)}_{11}(s)
-J_1(s)[t^{(2)}_{11}(s)]^2}\delta_{i1}\delta_{j1},
\eeq
and subtract it  from $T_{11}$. The extracted $f_0$ amplitude is given 
as~\cite{DalitzTuan}
\beq
f_0(s)=(T_{11}(s)-U_{11}(s))e^{-2i\delta_U(s)}
\eeq
where $\delta_U$ is the phase shift of $U_{11}$.  From this 
definition the mass and the width of the $f_0$ state are 
\beq
M_{f_0}=977.7 ~{\rm MeV}\qquad \Gamma_{f_0}=35.6~{\rm MeV},
\label{Upole}
\eeq
where $M_{f_0}$ is defined as a mass at which the phase shift of $f_0(s)$ 
crosses $90^\circ$, and  
\beq
\Gamma_{f_0}^{-1}=\frac{1}{2}
\frac{d\delta_{f_0}(s)}{d\sqrt{s}}|_{\sqrt{s}=M_{f_0}}.
\eeq
The values in Eq.(\ref{Upole}) are very close to the pole position in  
Table~\ref{tab:pole}. 
In this case there are no background term in $\pi\pi\to K\bar K$ and 
$K\bar K\to K\bar K$ amplitudes, and $T_{12}$ and $T_{22}$ are 
dominated by the $f_0$ state alone. Indeed, the clear peak structure 
appears both in these amplitudes, but it would not be true that there 
are no background term in these amplitudes.

\section{The $f_0$ and $a_0$ states in production processes}
\subsection{Radiative $\phi$ meson decays}
Many theoretical calculations of the radiative $\phi$ meson decays    
are based on the common production mechanism, in which the final S-wave 
two-meson scattering amplitude is connected to the initial decay vertex 
through the charged kaon loop. The validity of the on-shell 
factorizability of the loop and the scattering amplitudes is discussed 
in Refs.~\cite{OOBS,Oller1,Marko,Nieves}.

According to Ref.\cite{Marko} we write the decay amplitude of the 
$\phi$ meson to two-meson state $\gamma m_1m_2$ as
\beq
F(\phi\to\gamma m_1m_2)=
2eg \{ G_K(s)+f\frac{m_\phi^2-s}{2m_\phi^2}J_K(s) \}
T_{K^+K^-\to m_1m_2}, \label{decayamp}
\eeq
where $m_1m_2$ means $\pi^0\pi^0$, $\pi^0\eta$ or $K^0\bar K^0$, and 
$g$ and $f$ are the parameters defined as 
\beq
g= \frac{G_Vm_\phi}{\sqrt{2}f_\pi^2}\quad\mbox{ and }\quad 
f=\frac{F_V}{2G_V}-1
\eeq
with $G_V$ and $F_V$ being the constants in the chiral 
Lagrangian\cite{Ecker,Huber,Marko} and a factor $1/\sqrt{2}$ in $g$ 
is the ratio of the coupling constant $\rho\pi\pi$ to $\phi K\bar K$. 
The triangle $K\bar K$ loop integral $G_K(s)$, which connects the 
$\phi$ meson to the S-wave two-meson state after emitting a 
photon, is given as~\cite{AchaIva,Bramon0,Kumano} 
\beqa
G_K(s)&=&\frac{1}{(4\pi)^2}\left\{1+\frac{m_k^2}{s-m_\phi^2}
\left[\log^2\left(\frac{\sigma_K(s)+1}{\sigma_K(s)-1}\right)-
\log^2\left(\frac{\sigma_K(m_\phi^2)+1}{\sigma_K(m_\phi^2)-1}
\right)\right] \right. \\ \noeq
&&\left.-\frac{m_\phi^2}{s-m_\phi^2}\left[\sigma_K(s)
\log\left(\frac{\sigma_K(s)+1}{\sigma_K(s)-1}\right)
-\sigma_K(m_\phi^2)\log\left(\frac{\sigma_K(m_\phi^2)+1}
{\sigma_K(m_\phi^2)-1}\right)\right] \right\},
\eeqa
and $J_K(s)$ is the kaon two-point loop function given in 
Eq.(\ref{Loop}). 
The scattering amplitude $T_{K^+K^-\to m_1m_2}$ is
$T_{K^+K^-\to \pi\pi}$ or $T_{K^+K^-\to \pi\eta}$, which is obtained 
in the previous section.
We adopt the values $G_V=55$ MeV and $F_V=165$ MeV given in \cite{Marko}, 
which are suited to the $\phi\to K^+K^-$ and $\phi\to e^+e^-$ decay 
widths, respectively, and then we have $g=4.69$ and $f=0.5$. 
We emphasize that there are left no adjustable parameters, and 
then the mass spectra are a proving ground of the validity of 
$T_{12}(s)$ below the $K\bar K$ threshold.

The mass dependence of the two-meson state in the radiative decay 
is given as 
\beq
\frac{d\Gamma(s)}{d\sqrt{s}}=\left(\frac{\alpha}{3\pi}\right)
\left(\frac{g^2}{4\pi}\right)\frac{k_f(m_\phi^2-s)}{m_\phi^3}
\left|\tilde F(\phi\to\gamma m_1m_2)\right|^2,
\eeq
where $k_f$ is the momentum of a final meson in the rest frame 
of the final $(m_1m_2)$ state, 
and we define $\tilde F$ as $F=2eg\tilde F$. The $s$-dependence of the 
phase space factor $k_f(m_\phi^2-s)/m_\phi^3$ has 
the maximum  near the middle of the whole $\sqrt{s}$-range, and 
vanishes at both ends. Furthermore, the loop integral  
$G_K(s)+f(m_\phi^2-w^2)/(2m_\phi^2)\cdot J_K(s)$ has a strong cusp 
behavior  at the $K\bar K$ threshold, which is very near to $m_\phi$, 
so that the mass distributions near 1 GeV are affected double by 
these kinematical factors. 
\begin{figure}[h!]
\begin{center}
\epsfxsize=12cm
\centerline{\epsfbox{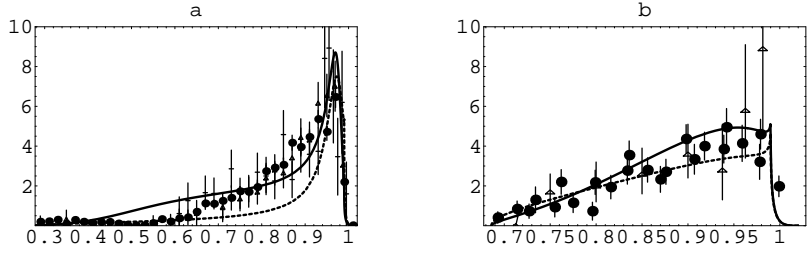}}
 \caption{(a) $dB(\phi\to\pi^0\pi^0)/d\sqrt{s}\times 10^4/{\rm GeV}$. 
 The solid line is the result of the full amplitude and the dotted one 
 that of the pole-terms. Experimental data are taken 
Experimental data are taken from \cite{CMD2,SNDpipi,KLOEpipi}. 
(b) $dB(\phi\to\gamma\pi^0\eta)/d\sqrt{s}\times 10^4/{\rm GeV}$. The 
lines are the same as in (a). Experimental data are taken 
 from \cite{SNDpieta, KLOEpieta}.}
\label{fig:phipipieta}
 \end{center}
 \end{figure}

The calculated mass distributions of the $\phi\to\gamma\pi^0\pi^0$ and 
$\phi\to\gamma\pi^0\eta$ decays reproduce the experimental data fairly 
well except for the mass region from 400 to 650 MeV of the 
$\phi\to \gamma\pi\pi$ decay, where the mass 
distribution shows a shallow dip.  See Fig.~\ref{fig:phipipieta}. 
Our results strongly indicate the validity of both of our scattering 
amplitudes and the production mechanism. The calculated and experimental  
branching fractions are tabulated in Table~\ref{tab:theory} and 
\ref{tab:branchratio}, respectively. 
The $\pi\pi$ mass distribution below 700 MeV for 
$\phi\to\gamma\pi^0\pi^0$ is affected by dominant backgrounds 
$\omega\pi^0\to \gamma\pi^0\pi^0$ and $\rho\pi^0\to\gamma\pi^0\pi^0$ 
as stressed in \cite{CMD2,SNDpipi}. It is better, therefore, to compare 
a theoretical calculation with experimental date above 700 MeV. 

\begin{table}[h]
\begin{center}
\begin{tabular}{|c|c|c|}\hline
&$B(\phi\to\gamma\pi^0\pi^0)$&$B(\phi\to\gamma\pi^0\eta)$\\ \hline
Whole mass range&$1.29\times 10^{-4}$&$0.875\times 10^{-4}$\\ \hline
$w~>~700$ MeV& $0.917\times 10^{-4}$&\\ \hline
$w~>~900$ MeV& $0.493\times 10^{-4}$&\\ \hline
Pole contribution&$ 0.655\times 10^{-4}$& $0.701\times 10^{-4}$\\ \hline
\end{tabular}
\caption{Branching fractions integrated over the indicated mass ranges.}
\label{tab:theory}
\end{center}
\end{table}
\begin{table}[h]
\begin{center}
\renewcommand{\arraystretch}{1.2}
\begin{tabular}{|c|c|c|c|}\hline 
$B(\phi\to\gamma\pi^0\pi^0)\cdot10^4$&CMD-2&SND&KLOE\\\hline
whole mass range&$1.08\pm 0.17\pm 0.09$&
$1.158\pm 0.093\pm 0.052$&$1.09\pm 0.03\pm 0.05$\\ \hline
$w~>~700$ MeV&$0.92\pm 0.08\pm 0.06$&$1.034\pm 0.066\pm 0.046$&
$0.96\pm 0.02\pm 0.04$\\\hline
$w~>900$ MeV&$0.57\pm 0.06\pm 0.04$&$0.559\pm 0.053\pm 0.025$&
\\ \hline\hline
$B(\phi\to\gamma\pi^0\eta)\cdot10^4$&CMD-2&SND&KLOE\\ \hline
whole mass range&$0.90\pm 0.24\pm 0.10$&$0.88\pm 0.14\pm 0.09$&
$0.851\pm0.051\pm 0.057^*$\\ \hline
\end{tabular}
\caption{Experimental data on the branching fractions, where CMD-2 
is referred to \cite{CMD2}, SND to \cite{SNDpipi,SNDpieta} and KLOE 
to \cite{KLOEpipi,KLOEpieta}. $^*$ means the data in which $\eta$ is 
identified through $\eta\to \gamma\gamma$, and it becomes 
$(0.796\pm 0.060\pm 0.040)$ when $\eta\to3\pi$ decay is used.} 
\label{tab:branchratio}
\end{center}
\end{table}
Our results $B(\phi\to\gamma\pi^0\pi^0)=0.917\times 10^{-4}$ above 
700 MeV and $B(\phi\to\gamma\pi^0\eta)=0.875\times 10^{-4}$ are 
consistent with the data in Table~\ref{tab:branchratio}. 
The problem is what parts of the mass distributions can be attributed 
to the contributions from "pure" $f_0(980)$ and $a_0(980)$ states. 
If we use the pole terms, the usual Breit-Wigner form for $f_0$ and 
the Flatt\'e form for $a_0$ given the previous section, we have 
$B(\phi\to\gamma f_0\to\gamma\pi\pi)=(0.655\times 3=1.965)\times 10^{-4}$ 
and $B(\phi\to\gamma a_0\to\gamma\pi\eta)=0.701\times 10^{-4}$, both 
of which are the values integrated over the whole mass range. The $f_0$ 
pole contribution above 900 MeV is $0.413 \times 10^{-4}$, that is 
84~\% of the $\pi^0\pi^0$ distribution in the same mass range,  and 
$B(\phi\to\gamma a_0\to\gamma\pi\eta)=0.701\times 10^{-4}$ is 80~\% of 
the whole $\pi\eta$ mass distribution. The pole contribution dominates 
the mass distributions above 900 MeV for the $\gamma\pi^0\pi^0$ decay, 
while the pole contribution dominates the whole mass distribution for 
the $\gamma\pi^0\eta$ decay.  Here we note that these 
values can be regarded as the branching fractions of the whole 
$B(\phi\to\gamma f_0)$ and $B(\phi\to\gamma a_0)$,  because the $K\bar K$ 
contributions are negligibly small due to the small phase space.

Finally we obtain    
\beq
B(\phi\to\gamma K^0\bar K^0)=4.26\times 10^{-8} \label{KbarK}
\eeq
for the neutral kaon pair decay. In order to get this value the mass 
difference between the neutral and charged kaon plays an important role 
to suppress the phase space; we use $497.67$ MeV for $K^0$ here, while 
$495$ MeV is used for the kaon mass throughout this paper. Our result is 
very similar to the values of previous works; $4.36\times 10^{-8}$ in the 
second paper of \cite{AchaGub} and $5.0\times 10^{-8}$ in  \cite{Oller1}. 
The charged kaon pair production is dominated by the Born terms 
including the Bremsstrahlung terms, and then it is almost irrelevant to 
the $f_0$ and $a_0$ states.

\subsection{Two-photon collision processes}
Two-photon collision processes are not so suitable to 
study the $f_0$ states in contrast to the radiative $\phi$ meson decays,  
because the charged kaon loop dominance does not hold. The Born terms 
with the charged pion exchange and $\omega$-meson exchange contribute 
to $\pi^+\pi^-$ and $\pi^0\pi^0$ production, respectively, and the 
charged pion loop and $\omega\pi^0\pi^0$ loop diagrams contribute to 
these production processes. And there are contributions from higher 
partial waves and the phase space are opened. Indeed, experimental 
$\pi\pi$ mass distributions seem not to show manifest signals of 
the $f_0$ state\cite{Crystalpi0pi0, Boyerpipm,CELLO}. 
On the other hand the $a_0$ production seems not to be worried with 
various production mechanisms besides the charged kaon triangle loop 
diagram as in the radiative $\phi$ meson decay. We discard the 
$\omega\gamma\eta$ and $\rho\gamma\pi$ vertices, because their coupling 
constants are much smaller than that of the $\omega\gamma\pi^0$ vertex.   
Focusing on the $S$-wave pion pair and $\pi^0\eta$ production processes 
we study $\gamma\gamma \to m_1m_2$ processes through the 
charged pion and kaon loop diagrams and the $\omega$ exchange diagrams, 
which are developed in Refs.\cite{Mennessier,DonHol,OOgamma}.

The production amplitude $F_f(w)$ from the initial photon state with the 
helicity $(++)$ to the final $S$-wave $m_1m_2$ state $f$ can be written 
\beq
F_f(s)=-2e^2\left(B_f(s)+\sum_{i=1,3}T_{fi}(s)G_i(s)\right), \label{ProdAmp}
\eeq
where the final state $f$ denotes charged pion pair, neutral pion pair and 
$\pi^0\eta$ states, and the intermediate state $i=1$ means the charged pair, 
$i=2$ the charged kaon pair state, and $i=3$ the $\pi^0\pi^0$ 
state generated in the $\omega\pi^0\pi^0$ triangle loop.  
$B_f(s)$ is the Born term, but $B_{\pi\eta}=0$, and $T_{fi}$ is the 
$i\to f$ scattering amplitude.  
The triangle loop functions $G_1(w)$ and $G_2(s)$ are the 
integrals of the charged pion and kaon loop diagram, respectively, 
which is given as 
\beq
G_i(s)=\frac{1}{(4\pi)^2}\left\{1+\frac{m_i^2}{s}\log^2\left(
\frac{\sigma_i(s)+1}{\sigma_i(s)-1}\right)\right\}. \label{triangle}
\eeq
The loop function $G_3$ with the $\omega$ exchange without any form 
factor is given in Ref.\cite{Mennessier}, which is rewritten in 
Appendix because of a lengthy expression. 
The discontinuity of $G_i(s)$ across the physical 
cut gives the Born term as 
\beq
{\rm Im}G_i(s)=-\rho_i(s)B_i(s)=-\rho_i(s)\left[
\frac{2m_i^2}{s\sigma_i(s)}
\log\left(\frac{1+\sigma_i(s)}{1-\sigma_i(s)}\right)\right] 
\eeq
for $i=1$ and $2$, and 
\beq
{\rm Im}G_3(s)=-\rho_1(s)B_3(s)=-\rho_1(s)\cdot\frac{-R_\omega}{\sqrt{2}}
\left\{-s+\frac{M_\omega^2}{\sigma_1(s)}\log\left(
\frac{1+v(s)+\sigma_1(s)}{1+v(s)-\sigma_1(s)}\right)\right\},
\eeq
where $B_3$ is the $\omega$ exchange Born term with 
$R_\omega=1.35~{\rm GeV}^{-2}$~\cite{DonHol}, 
$v(s)=2(M_\omega^2-m_\pi^2)/s$, and a factor $1/\sqrt{2}$ is related to 
the definition of the cross section. 
The production amplitude $F_f$ satisfies the final interaction theorem as   
\beq
 {\rm Im}F_f(s)=-\sum_iT_{fi}^*(s)\rho_i(s) F_i(s) ,\label{FSI}. 
\eeq
The S-wave two-meson production cross section of
the $f$ state is written as 
\beq
 \sigma^f_S(s)=2\pi\alpha^2\frac{\sigma_f(s)}{s}
 \left|\tilde F_f(s)\right|^2
\eeq
with $F_f=-2e^2\tilde F_f$, and $k_f$ is the CM momentum. 
 \begin{figure}[h!]
\begin{center}
 \epsfxsize=16cm
 \centerline{\epsfbox{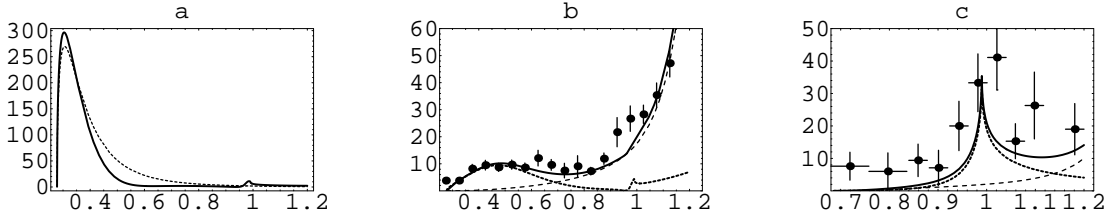}}
 \caption{ (a) $\gamma\gamma\to\pi^+\pi^-$ cross section integrated 
 over $|\cos\theta|\le 0.6$. The solid line is the total $S$-wave 
 cross section, and the dashed one the Born term. The behavior near 
 the $f_0$ region is given in the inner box. 
 (b) $\gamma\gamma\to\pi^0\pi^0$ cross section integrated over 
 $|\cos\theta|\le 0.8$. The solid line shows the sum of the $S$-wave   
 cross section and the $f_2(1280)$ resonance. The dotted line shows 
 the $S$-wave cross section and the dashed one the $f_2(1280)$ cross 
 section. The experimental data are taken from \cite{Crystalpi0pi0}.
 (c) $\gamma\gamma\to\pi^0\eta$  cross section integrated over 
 $|\cos\theta|\,<\, 0.9$ with the $a_2(1320)$ resonance.  The lines are 
 the same as in (b). The experimental data are taken from  
 \cite{Crystalpieta,JADEpieta}}
\label{fig:photoMM}
\end{center}
\end{figure}

The calculated $S$-wave $\pi^+\pi^-$ cross section integrated over the 
region $|\cos\theta|<0.6$ shows a large peak resulting from the Born 
term and a small $f_0$ signal coming from the rescattering terms as 
shown in Fig.~\ref{fig:photoMM}~(a). The cross section integrated over 
the full angle is merely 18.6 nb/GeV at the $f_0$ peak.  
This is due to destructive interferences among the Born term and various 
rescattering terms.  The small $f_0$ signal 
in the charged pion distribution appears to be consistent with the 
experimental mass distributions, which show no clear $f_0$ 
signal\cite{Boyerpipm,CELLO}. 
The calculated $S$-wave $\pi^0\pi^0$ cross section integrated over the 
region $|\cos\theta|\,<\,0.8$ shows a dip and a cusp, but not the $f_0$ 
signal. If we discard both the $\omega$ exchange Born term and the 
rescattering term with the $\omega\pi\pi$ loop, we can see a clear 
$f_0$ peak. This implies that the present form of the $\omega$ exchange  
amplitudes without any form factor strongly reduces the $f_0$ signal. 
We should remember, here, that the $T_{11}$ amplitude does 
not show a peak but a sharp dip in the $f_0$ region, and that the 
rescattering terms with the charged pion loop and the $\omega\pi\pi$ 
loop contain the very $T_{11}$ amplitude. The broad bump around 500 
Mev comes mainly from the pion loop diagram with $T_{11}$. 
Our neutral pion mass distribution seems not to be inconsistent with 
the experimental mass distribution, but there may be some deficiencies 
in the region of $f_0$ state. We think, however, 
that these possible deficiencies do not imply the invalidity of the OOP 
amplitudes, but suggest the inadequacy of the treatment of production 
mechanisms.

As to the $\pi^0\eta$ production channel the sharp peak appears in our 
calculation, but the consistency with the experimental data is not clear. 
It is noted that we take into account the rescattering term with the 
charged kaon loop alone in order to calculated the $\pi^0\eta$ mass 
distribution. If there contribute other production mechanisms, the 
similar shape of the scattering amplitudes would not destroy the sharp 
peak structure. As stressed in Ref.~\cite{OOgamma} the axial vector meson 
exchange would enhance the peak.

The partial widths of the $f_0(980)\to\gamma\gamma$ and  
$a_0(980)\to\gamma\gamma$ have been expected to play a role to  
distinguish the structure of them as in the radiative $\phi$  
decays\cite{AchaShes}.   
We have learned in the above that different production mechanisms and 
their interference mask the $f_0$ signal in the charged and neutral pion 
pair production. 
We should be cautious, therefore, of the way how to estimate the partial 
$f_0\gamma\gamma$ width.  

Taking the narrow width limit of the relativistic Breit-Wigner formula 
given in \cite{Crystalpi0pi0,Boyerpipm}, we can define the partial width 
by integrating the mass spectrum around the peak region as 
\beq
\Gamma_{\gamma\gamma}( R\to  f)=\frac{M_R }{8\pi^2}
\int_{s_1}^{s_2} ds\sigma_{\gamma\gamma\to f}(s),
\eeq
where $\sqrt{s_1}~(\sqrt{s_2})$ is $0.9~(1.05)~{\rm GeV}$ for $f_0$ and  
$0.8~(1.2)~{\rm GeV}$ for $a_0$. The integration range is chosen so as 
reflect the form of the mass spectra. We obtain  
\beq
\Gamma_{\gamma\gamma}(f_0\to\pi^+\pi^-)=0.056~{\rm keV}, 
\eeq
but we cannot estimate $\Gamma^{\gamma\gamma}(f_0\to\pi^0\pi^0)$, 
because there is no $f_0$ signal. If we use the pole terms, $P_{ij}(s)$,  
written by the Breit-Wigner forms instead of $T_{ij}$, we 
observe peaks in both spectra deformed by the cusp behavior, and 
we obtain  
\beq\begin{array}{rcl}
 \Gamma_{\gamma\gamma}(f_0\to\pi^+\pi^-)&=&0.081~{\rm keV},\\
 \Gamma_{\gamma\gamma}(f_0\to\pi^0\pi^0)&=&0.035~{\rm keV}.
\end{array}
\eeq
The sum is $0.116$ keV, and we have
\beq 
\Gamma_{\gamma\gamma}(f_0\to~{\rm all})=0.20~{\rm keV},
\eeq
if we use $B(f_0\to \pi\pi)=0.58$ calculated from the partial widths 
Eq.(\ref{f0pole}). 

Though the $\pi\eta$ spectrum cannot be expressed by the Breit-Wigner 
form, we use the same equation with the peak at the $K\bar K$ 
threshold, 990 MeV, and obtain   
\beq
\Gamma_{\gamma\gamma}(a_0\to\pi\eta)=0.17~{\rm keV}. 
\eeq
If we use $(-{\rm Im}T_{22}\cdot|G_2(s)|^2)$ instead of 
$(\rho_1(s)|T_{12}G_2(s)|^2)$ in order to estimate the total 
photo-partial width\cite{OOgamma,Oller2}, where $T_{22}$ is 
$T_{K^+K^-\to K^+K^-}$ and $T_{12}$ is $T_{K^+K^-\to\pi^0\eta}$, 
we obtain a value, 
\beq
\Gamma_{\gamma\gamma}(a_0\to~{\rm all})=0.27~{\rm keV}.\label{alla0}
\eeq 
Due to the similarity of the pole terms to the full amplitudes, 
the photo-partial widths are almost unchanged.  

Preceding theoretical estimates are 
\beqa
f_0(980)&&\left\{\begin{array}{rclr}
\Gamma_{\gamma\gamma}(f_0\to~{\rm all})&=&0.20~{\rm keV} &
\cite{OOgamma},\\
\Gamma_{\gamma\gamma}(f_0\to~{\rm all})&=&0.28^{+0.09}_{-0.13}~{\rm keV}&
 \cite{BogPengamma}. 
 \end{array}\right.\\
 a_0(980)&&\left\{\begin{array}{rclr}
\Gamma_{\gamma\gamma}(a_0\to~{\rm all})&=&0.27~{\rm keV} &
\cite{AchaShes},\\
\Gamma_{\gamma\gamma}(a_0\to~{\rm all})&=&0.78~{\rm keV} &
\cite{OOgamma}.
\end{array}\right.
\eeqa

The experimental data using the Breit-Wigner fits are summarized as follows: 
\beqa
f_0(980)&&\left\{\begin{array}{rclr}
\Gamma_{\gamma\gamma}(f_0\to~{\rm all})&=&0.29\pm 0.07\pm 
0.12~{\rm keV} &\cite{Boyerpipm},\\
\Gamma_{\gamma\gamma}(f_0\to~{\rm all})&=&0.31\pm 0.17~{\rm keV} &
\cite{Crystalpi0pi0},
\end{array}\right.\\
a_0(980)&&\left\{\begin{array}{rclr}
\Gamma_{\gamma\gamma}(a_0\to\pi\eta)&=&0.19^{+ 0.12}_{-0.10}~{\rm keV} &
\cite{Crystalpieta},\\
\Gamma^{\gamma\gamma}(a_0\to\pi\eta) &=&0.28\pm 0.04\pm 0.10~{\rm keV} 
&\cite{JADEpieta},\\
.\Gamma_{\gamma\gamma}(a_0\to\pi\eta) &=&0.24\pm 0.08~{\rm keV} 
&\cite{Amsler}.
\end{array}\right. 
\eeqa
In the above $\Gamma_{\gamma\gamma}(f_0\to~{\rm all})$ of 
\cite{Boyerpipm} and \cite{Crystalpi0pi0} are obtained 
from $\pi^+\pi^-$ and $\pi^0\pi^0$ spectrum, respectively, using 
the conventional isospin factor and the old PDG value of 
$B(f_0\to\pi\pi)=0.78$\cite{OldPDG}. 
It is dubious, however, to use a conventional isospin factor to estimate 
the full $\pi\pi$ decay width, because complicated interferences should 
probably mask the $f_0$ signal.  
As to the $a_0$ signal our value 0.17 keV is not inconsistent with the 
value 0.19 keV of \cite{Crystalpieta}, but is a little bit smaller than 
the average value\cite{Amsler}.

Finally we briefly comment on $J/\psi$ decays. 
The decays $J/\psi\to (\phi~{\rm or}~\omega)(\pi\pi)$ and 
$J/\psi\to \rho(\pi\eta)$ are also expected to specify the quark 
contents of the $f_0$ and $a_0$ states\cite{AchasovKEK}. The two-pion 
mass spectrum of the former decay accompanied by $\phi$ shows a clear 
peak at the $K\bar K$ threshold as like as in 
$\phi\to\gamma(\pi\pi)$\cite{CMD2JPhi}. 
This strongly suggests that the decay proceeds mainly through the kaon 
loop so as to make $T_{K\bar K\to\pi\pi}$ work effectively.
On the other hand the mass spectrum of the two-pion decay with $\omega$ 
shows a large broad peak centered at about 450 MeV and a dip-bump 
structure near the $K\bar K$ threshold\cite{CMD2JOmeg}, that 
resembles to the $\gamma\gamma\to\pi^+\pi^-$ process. The $f_0$ 
signal seems to be masked. Indeed, the mass spectrum  
calculated in terms of the Born term, that is a direct $S$-wave $\pi\pi$ 
production term, the pion and kaon loop terms and a sequential decay 
mode $b_1(1240)\pi\to\omega\pi\pi$ gives the broad peak originated from  
the Born term and a dip-cusp structure instead of the $f_0$
 peak\cite{MUptp}.\footnote{Recalculation with the present 
 set of the LECS also shows a dip with a rather small cusp.} 
Further higher spin meson exchange diagrams would not be discarded.
 
The $J/\psi\to\rho(\pi\eta)$ decay would also include some 
decay mechanisms similar to the $J/\psi\to\omega(\pi\pi)$ decay, but 
the $a_0$ signal might not be masked so much as contrasted with the 
$f_0$ case, since the isovector scattering amplitudes have similar shape  
and do not have large backgrounds.

\section{Large $N_c$ behavior of scalar scattering amplitudes}
In the previous two sections we have found that the OOP version of 
unitarized chiral theories can reproduce not only scattering but 
also production processes fairly well.   
It is not clear, however, what this success tells us about the 
so-called quark contents of the $f_0(980)$ and $a_0(980)$ states, 
because the OOP amplitudes contain phenomenological constants, which 
should be determined so as to be suited to the experimental data. 

It is said that ChPT is an effective theory of low energy 
QCD\cite{WittenChiral, GL1}, and $q\bar q$ meson states become 
weakly interacting narrow resonances in the large $N_c$ limit of 
QCD\cite{tHooft,Witten}.  
It offers a useful means to reveal the the quark contents, therefore,
to examine whether the scalar mesons survive as narrow resonances  
when $N_c$ increases from the physical value $N_c=3$.  
(See Ref.~\cite{MU0801} for details.)

Because the pion decay constant $f_\pi$ is of 
$O(N_c^{1/2})$ and the LECs, $L_1$, $L_2$, $L_3$, $L_5$ and $L_8$ 
are to be of $O(N_c)$, but $2L_1-L_2$,  $L_4$, $L_6$ and $L_7$ are 
of $O(1)$~\cite{GL2, PerisRaf, Siklody,Peris}, we put
\beqa
L_n(N_c)&=& {\widehat{L_n}}\cdot\frac{N_c}{3}+\Delta L_n, \\
f_\pi^2(N_c)&=&\widehat{f_\pi^2}\cdot\frac{N_c}{3},
\eeqa
where $\widehat{L_n}$ satisfy the relations, $2\widehat{L_1}-
\widehat{L_2}=\widehat{L_4}=\widehat{L_6}=\widehat{L_7}=0$, and 
$\Delta L_n$ are of $O(1)$. Thus, we have 
\begin{equation}
\frac{L_n}{f_\pi^2}=\frac{\widehat{L_n}}{\widehat{f_\pi^2}}
+\frac{\Delta L_n}{\widehat{f_\pi^2}}\cdot\frac{3}{N_c} 
\label{Nc}
\end{equation}
with $\widehat{f_\pi}=93$ MeV.  In the case of our LECs $\Delta L_n$ 
are given as 
\beq
\Delta L_2\times 10^3=-0.10,\mbox{ and }\Delta L_7\times 10^3=-0.25,
\eeq
and others are zero. The $T^{(2)}(s,t,u)$ 
is of $O(N_c^{-1})$ and $T^{(4)}_{\rm poly}(s,t,u)$ is also of 
$O(N_c^{-1})$, because $L_n/f_\pi^2$ scales as $O(N_c^0)$ as seen in 
Eq.(\ref{Nc}).  The $s$-channel loop term given by 
$t^{(2)}(s)J(s)t^{(2)}(s)$ is of $O(N_c^{-2})$.
This difference of the $N_c$ dependence produces different behavior 
of the amplitudes when $N_c$ becomes large. The change of the 
renormalization scale affects values of $\Delta L_n$, but we do not 
consider the scale change explicitly because the $\Delta L_n$ terms 
fade out as $N_c$ increases.

At first, we discuss the behavior of the $\rho$ meson in the single 
channel calculation. The mass of the $\rho$ meson is given as a zero 
of a sum of the term $[t^{(2)}-t^{(4)}_{\rm poly}]$  of $O(N_c^{-1})$ 
and the real part of the loop contribution of $O(N_c^{-2})$. Since the 
mass is controlled by a combination of the LECs $(2L_1-L_2+L_3)$ of 
the former term~\cite{DP}, the mass stays almost constant for 
$N_c\geq 3$. Because the imaginary part of the loop term contributes 
to the  width, the width decreases as $O(N_c^{-1})$. Thus, the 
$\rho$ meson remains as narrower resonance almost at the same point as 
$N_c$ increases as shown in Fig.~{\ref{fig:rho}}. Other vector mesons 
such as $K^*(890)$ behave as the same as the $\rho$ meson. If we 
extend the calculation to the two-channel OOP version, the results 
do not change.  Thus, we can conclude that the vector mesons 
described by the OOP version behave as narrow resonances, and then 
have the nature consistent with the $q\bar q$ mesons. 
\begin{center}
\begin{figure}[h]
\epsfxsize=10 cm
\centerline{\epsfbox{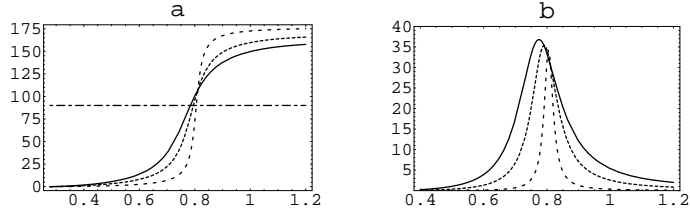}}
\caption{$N_c$ dependence of phase shift (left) and 
 cross section (right) of the $\rho$ channel. Lines correspond to 
 $N_c=3$, 5 and 15 from the top to the bottom.}
 \label{fig:rho}
\end{figure} 
\end{center}

As to the scalar mesons the situation is drastically changed. 
The $N_c$ dependence of the phase shift and the cross section of the 
$(I,~J)=(0,~0)$ channel are shown in Fig.~{\ref{fig:LN00}}, where 
$N_c$ increases from 3 to 12. In contrast to the vector channel we 
observe that the phase shift becomes flat and the cross section fades 
out as $N_c$ becomes large; the sharp rise of the phase shift near 
the $K\bar K$ threshold and the large bump of the cross section near 
500 MeV seen at $N_c=3$  to 5 disappears even at $N_c=6$, 
and then the phase shift and the cross section 
become almost flat and fade out. Similar drastic change in the $N_c$ 
dependence has also been observed in Ref.~\cite{Harada}, though it 
is in quite different context. 
\begin{center}
\begin{figure}[h]
\epsfxsize=10 cm
\centerline{\epsfbox{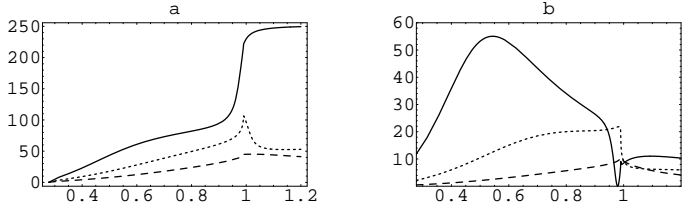}}
\caption{$N_c$ dependence of the phase shift~(left) and the 
 cross section~(right) of the (0,0) channel. Solid, dotted, 
 and dashed lines correspond to $Nc=$3, 6 and 12 cases, respectively.}
 \label{fig:LN00}
 \end{figure}
 \end{center}

The $(I,~J)=(1,~0)$ channel contains the $a_0(980)$ state and 
appears as a cusp-like sharp peak at $N_c=3$ as seen in 
Fig.~{\ref{fig:LN10}}.
The rising phase shift after the cusp bends down and becomes to a 
flat curve, and the cross section having  a sharp peak fades out as 
$N_c$ increases. The $(I,~J)=(1/2,~0)$ channel show the same behavior 
as $(0,~0)$ and $(1,~0)$ channels. 
\begin{center}
\begin{figure}[h]
\epsfxsize=10 cm
\centerline{\epsfbox{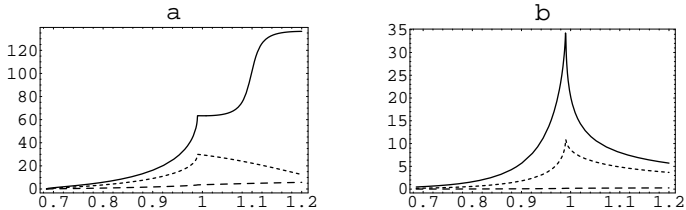}}
\caption{$N_c$ dependence of the phase shift~(left) and the
 cross section~(right) of the (1,0) channel. $N_c=3$, 4 and 12 from 
 the top to bottom.}
 \label{fig:LN10}
\end{figure} 
\end{center}

We have shown in Sec.~{\bf 3} that the $f_0(980)$ pole exists at 
$(978.4-19.7~i)$ MeV and the $a_0$ pole at $(1110.4-8.9~i)$ MeV 
at $N_c=3$. Where do the poles go away as $N_c$ increases ? 
We make an approximate calculation of the pole positions by expanding 
the amplitudes in powers of $k_2$ up to the first order, where $k_2$ 
is the momentum of the $K\bar K$ channel. We observe that the poles 
move into the upper half plane of the IV sheet from the lower half 
plane of the II sheet, winding around the branch point at the 
$K\bar K$ threshold, and go away from the real axis as shown in 
Fig.~{\ref{fig:pole}}. This approximation gives 
$W_{f_0}=(977.9-18.5~i)$ MeV and $W_{a_0}=(1095.7+2.7~i)$ GeV at 
$N_c=3$, which are close to the above values. Pole positions at 
larger $N_c$ cannot be reliable owing to the rough approximation, 
but this behavior would remain unchanged. 
\begin{center}
\begin{figure}[h]
\epsfxsize=10 cm
\centerline{\epsfbox{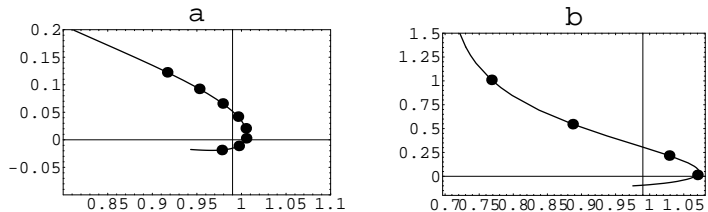}}
\caption{$N_c$ dependence of the $f_0(980)$ pole~(left) and 
$a_0(980)$ pole~(right). Both of the poles wind around the branch 
point at $K\bar K$ threshold to go upward on the IV sheet. Black point 
starts at $N_c=3$ and increases by 1 to 10 for $f_0$ and to 6 for $a_0$.
}
 \label{fig:pole}
\end{figure} 
\end{center}

We have calculated the $N_c$ dependence of the vector and scalar 
channels stating from $N_c=3$ to finite values, 30 for the vector 
channels and 15 for the scalar channels within the two-channel 
OOP version, and we have observed that the vector mesons survive as 
narrower resonances at almost the same position, but the resonant 
structures of the scalar channels fade out at rather low values of 
$N_c$ near 5 or 6. By extending these observations we are led to a 
conclusion that the light vector meson nonet has the nature 
consistent with the $q\bar q$ mesons in large $N_c$ QCD, but the 
light scalar meson nonet cannot survive in large $N_c$ and then 
cannot have the nature of the $q\bar q$ mesons. 
This conclusion is the same as obtained by Pel\'aez, excluding an 
exceptional case of the $a_0(980)$ state\cite{Pelaez}, where the 
full $O(p^4)$ amplitudes are used. 
It is also consistent with the result that the scalar mesons are 
describable without pre-existing tree resonance poles 
and then of dynamical origin, possibly except for 
$f_0(980)$\cite{OOND,Oller3}.  If one treats the $f_0$ state 
within the two-channel model, the pre-existing resonance pole is not 
needed\cite{Oller3}. Our conclusion supports the arguments that the 
light scalar mesons are of the $K\bar K$ 
molecule\cite{Wein,Lohse,Janssen}, and of $q^2{\bar q}^2$ 
states\cite{Jaffe,Jaffe4q,Achasov,AchaIva,AchaGub,Kumano}. It is 
interesting to see that the lattice QCD calculation suggests that 
the masses of scalar mesons with $I=0$ and $1/2$ are almost twice 
of those of the $\rho$ and $K^*(890)$ mesons, respectively, if they 
are composed of connected $q\bar q$ lines\cite{kunihiro}.

\section{Concluding remarks}
In this paper we have discussed $S$-wave scattering processes and 
radiative production processes of two-pseudoscalar mesons in a 
unified manner. The scattering amplitudes are constructed within 
the two-channel OOP version and satisfy exact $s$-channel unitarity. 
We emphasize that the results of the radiative $\phi$ meson decays 
are the prediction by the two-channel OOP scattering amplitudes, 
once the decays occur dominantly through the charged kaon loop.

We have found that the scattering data are reproduced rather well, 
though we do not attempt to get the best parameter set. Radiative 
production processes reveal the structure of the transition amplitudes 
$T_{12}$ below the $K\bar K$ threshold. Our results of the radiative 
$\phi$ meson decays show the validity of our amplitudes under the 
dominance of the $K\bar K$ loop mechanism as discussed in 
Sec.~{\rm 3.1}, where the $\pi^0\pi^0$ and $\pi^0\eta$ mass 
distributions are reproduced fairly well, at least above 700 MeV for 
the former channel, and then the branching fraction 
$B(\phi\to\gamma\pi^0\pi^0)$ above 700 MeV and 
$B(\phi\to\gamma\pi^0\eta)$ are close to the experimental data. 
 
We point out that the branching fraction $B(\phi\to\gamma f_0)$ 
is strongly model-dependent, though $B(\phi\to\gamma\pi^0\pi^0)$ 
integrated over a mass range given by different experimental groups 
are mutually consistent within errors as shown in 
Table~\ref{tab:branchratio}.  
The SND collaboration assumes that the whole mass distribution is 
dominantly given by the $f_0(980)$ resonance with 
$m_{f_0}=969.8\pm 4.5$ MeV and $\Gamma_{f_0\to\pi\pi}\approx 200$ 
MeV, and gives 
$B(\phi\to \gamma f_0)=(3.5\pm 0.3{}^{+1.3}_{-0.5})\times 
10^{-4}$~\cite{SNDpipi}. The CMD-2 collaboration  gives
$B(\phi\to\gamma f_0(980))=(3.05\pm 0.25\pm 0.72)\times 10^{-4}$
under the single $f_0(980)$ fit, but $(1.5\pm 0.5)\times 10^{-4}$  
under the two-resonance fit with $f_0(980)$ and 
$f_0(1200)$\cite{CMD2}. On the other hand KLOE collaboration fits 
the $\pi^0\pi^0$ mass distribution in terms of the 
$\gamma(f_0+\sigma(600))$ mode with 
$m_\sigma= 478$ MeV and $\Gamma_\sigma= 324$ MeV, which are obtained  
in \cite{Aitala}, and gives 
$B(\phi\to \gamma f_0)=3\times(1.49\pm 0.07)\times 10^{-4}
\cong 4.47\times 10^{-4} $~\cite{KLOEpipi,KLOEpieta}.
In this analysis the $\sigma$ contribution gives a bump near 500 
MeV, but the experimental mass distribution is very small there, 
so that the bump is erased by the destructive interference 
with the $f_0$ resonance with a large width, where $m_{f_0}=973$ 
MeV and $\Gamma(f_0\to\pi\pi)=260$ MeV. This enlarges the 
$\gamma f_0$ branching fraction. It is further pointed out that their 
amplitude violates unitarity\cite{BogPen}. We note that 
a careful analysis using K-matrix parametrization of $D_s^+$ and 
$D^+$ decays into three pions does not need such a low mass $\sigma$ 
pole in T-matrix\cite{FOCUS}.

In theoretical estimates performed by adjusting model parameters 
so as to fit the experimental data, the branching fraction over the 
whole mass range is almost consistent with each other. But 
theoretical estimates  of $B(\phi\to\gamma f_0)$ seem to be 
model-dependent similarly to the experimental estimates. Our 
integrated pole contribution $B(\phi\to\gamma f_0)=1.97\times 10^{-4}$ 
is much larger than the value $0.31\sim 0.34\times 10^{-4}$ estimated 
through calculating pole residues\cite{BogPen}, but smaller than the 
value $3.11\sim 3.19\times 10^{-4}$ obtained by integrating 
${\rm Im}T_{22}$ so as to estimate the 
$f_0$ pole contribution \cite{Oller2}. Thus, the branching fraction 
of the "pure" $f_0$ signal has not yet been established in 
theoretical works originating mainly from uncertainties of the 
parametrization of scattering amplitudes. 
On the other hand our result 
$B(\phi\to \gamma a_0)=0.7\times 10^{-4}$ is similar to values of 
other models.

Unfortunately, it is difficult to deduce a concrete result on 
the $f_0$ state from two-photon production processes, because 
the processes are not dominated by a single production mechanism, 
and the large charged pion exchange and $\omega$ exchange terms 
mask the $f_0$ signal originated from the charged kaon loop. 
Similar complex situation is seen in the 
$J/\psi\to\omega f_0$ process.  We have observed that mass spectra 
calculated by the pure pole terms, expressed by appropriate 
Breit-Wigner forms, are different from those calculated by full 
scattering amplitudes. 
This implies that the background coming from the $\sigma$ enhancement 
cannot be discarded in comparing calculated results with 
experimental data. 
 
It is impossible to conclude the quark contents of the light scalar 
mesons even if we succeed in describing scattering and production 
processes simultaneously in terms of the OOP amplitudes. We have 
proposed to study how scattering amplitudes behave as $N_c$ 
increases from 3, because ChPT is an effective theory of QCD and has 
the explicit $N_c$ dependence. As a result we have obtained the 
probable conclusion that while the light vector mesons are of typical 
$q\bar q$ mesons, all of the light scalar mesons below 1 GeV cannot 
survive as narrow resonances when $N_c$ increases from 3, and then  
the light scalar mesons cannot be of simple $q\bar q$ states.

If the mesons in the light scalar nonet are dominantly 
composed of hadronic or four quark component, but include 
$|q\bar q>_P$ with a small fraction as in Ref.~\cite{Close}, we 
could find out the small $|q\bar q>_P$ component by increasing $N_c$ 
in theoretical models, because the large hadronic or four quark 
component fades out\cite{Jaffe4q} and the $q\bar q$ 
component remains. At least, our calculation within the two-channel 
OOP approximation does not indicate that such an intriguing change 
will occur in larger $N_c$ region.  

Our conclusion strongly indicates that all of the light scalar 
mesons are dynamical effects originating from unitarity, chiral 
symmetry and channel couplings.

\appendix
\section{Isospin decomposition of scattering amplitudes and 
$\omega\pi^0\pi^0$ loop integral}
Here we summarize the isospin decomposition of amplitudes 
written in terms of charged states. 
\begin{eqnarray*}
<\pi^+\pi^-|T|\pi^+\pi^->&=&\frac{2}{3}T^0_{11}+\frac{1}{3}T^2,\\
<\pi^+\pi^-|T|\pi^0\pi^0>&=&\frac{\sqrt{2}}{3}(T^0_{11}-T^2),\\
<\pi^0\pi^0|T|\pi^0\pi^0>&=&\frac{1}{3}T^0_{11}+\frac{2}{3}T^2,\\
<K^+K^-|T|\pi^+\pi^->&=& \frac{1}{\sqrt{3}}T^0_{12},\\
<K^+K^-|T|\pi^0\pi^0>&=& \frac{1}{\sqrt{6}}T^0_{12},\\
<K^+K^-|T|\pi^0\eta>&=&\frac{1}{2}T^1_{12},\\
<K^+K^-|T|K^0\bar K^0>&=&\frac{1}{2}(T^0_{22}-T^1_{22}).
\end{eqnarray*}
These amplitudes satisfy unitarity relations from each other. \\

The $\omega\pi^0\pi^0$ triangle loop function is given in Appendix B 
of Ref. \cite{Mennessier}, which is rewritten so as to 
 fit our normalization and sign. 
\begin{eqnarray*}
G_3(s)&=&\frac{R_\omega}{\sqrt{2}(4\pi)^2}\left\{2M_\omega^2\tilde L_2
+s\sigma_1(s)\log\left(\frac{\sigma_1(s)+1}{\sigma_1(s)-1}\right)\right.
\\
 &-&\left. s\left(\frac{M_\omega^2-2m_\pi^2}{M_\omega^2-m_\pi^2}+
\left(\frac{M_\omega^2}{M_\omega^2-m_\pi^2}\right)^2\log
\frac{M_\omega^2}{m_\pi^2}\right)\right\},\\
\tilde L_2&=&-Sp(1+iu)-Sp(1-iu)-Sp(-\frac{1+\sigma_1}{v})-
Sp(-\frac{1-\sigma_1}{v})\\
&-&\log(1+\frac{1+\sigma_1}{v})\log(\frac{1+\sigma_1}{v})-
\log(1+\frac{1-\sigma_1}{v})\log(\frac{1-\sigma_1}{v})\\
&-&\frac{\pi^2}{3}+\pi\tan^{-1}u
+i\frac{\pi}{ 2}\log\left(\frac{1+v+\sigma_1}{1+v-\sigma_1}\right) 
\quad{\rm for}~s\,>\,2m_\pi^2,\\
v&=&\frac{2(M_\omega^2-m_\pi^2)}{s},\\
u&=&\frac{\sqrt{sM_\omega^2}}{M_\omega^2-m_\pi^2},
\end{eqnarray*}
where $Sp(z)$ is the double polylog function defined as 
$$ Sp(z)=-\int_0^z dx\frac{\log(1-x)}{x}. $$

\end{document}